C2SMART
CONNECTED CITIES WITH SMART TRANSPORTATION

WHITE PAPER ISSUE 3

# Agent-based Simulation Model and Deep Learning Techniques to Evaluate and Predict Transportation Trends around COVID-19

Ding Wang, Fan Zuo, Jingqin Gao, Yueshuai He, Zilin Bian, Suzana Duran Bernardes, Chaekuk Na, Jingxing Wang, John Petinos
Kaan Ozbay, Joseph Y.J. Chow, Shri Iyer, Hani Nassif, Xuegang Jeff Ban

Contact: c2smart@nyu.edu
c2smart.engineering.nyu.edu

## Executive Summary

The COVID-19 pandemic has affected travel behaviors and transportation system operations, and cities are grappling with what policies can be effective for a phased reopening shaped by social distancing. This edition of the white paper updates travel trends and highlights an agent-based simulation model's results to predict the impact of proposed phased reopening strategies. It also introduces a real-time video processing method to measure social distancing through cameras on city streets.

*Key Findings*

- New York City (NYC) traffic volumes are starting to increase, but average traffic speeds in the city remain high. Despite the increased vehicular volumes, no significant changes in average travel times were observed on NYC and Seattle corridors in May, compared with Apr 2020.
- Data from May 2020 shows a smaller drop of the subway ridership from 2019 levels compared to previous March and April drops from 2019, but more data is needed to verify the increase. Seattle's public transit ridership rebound continued to lag in May, indicating a lagging mode choice preference for public modes of travel.
- Weigh-in-motion (WIM) from C2SMART's testbed on the Brooklyn-Queens Expressway (BQE) showed the number of very heavy trucks (GVW > 100 kips) remain down 29% for Queens bound (QB) and 44% for Staten Island bound (SIB) traffic as of May 15 compare to Feb 2020.
- An agent-based simulation model (MATSim-NYC) was used to predict how this pandemic is changing travel behavior provide some insights for the reopening of NYC.
    - Because of changed preferences, a full reopening would perhaps only see as much as 73% of pre-COVID transit ridership and an increase in the number of car trips by as much as 142% of pre-pandemic levels, assuming mode preferences held during the crisis are maintained. Evidence from other cities further in the reopening process points to lingering mode choice preferences from the pandemic during reopening, however due to the unique mode-choice in NYC these numbers may represent an extreme case.
    - The effect of applying a capacity restriction on public transit, as is being applied in some cities, was also studied. Limiting transit capacity to 50% would decrease transit ridership to potentially as low as 64% while increasing car trips to as much as 143% of pre-pandemic levels by Phase 4 of NY's reopening plan.
- A deep-learning based video-processing algorithm was developed to measure social distancing using installed cameras in cities.
    - A gradual increase in pedestrian density has been observed in multiple locations in NYC in the last two weeks of May. It was observed that the average percentage of pedestrians following social distancing guidelines of 6 feet apart at select locations dropped slightly from 91% on Apr 2 to 86% on May 27.
    - At 5th Avenue and 42nd Street, while density observed from cameras remained low, peak hour spikes in car and pedestrian density began to emerge in May. Meanwhile, cyclist density has almost approached pre-pandemic levels at this location.

## Mobility Trends

Average subway ridership in NYC was down 75% and vehicular traffic via MTA bridges and tunnels was down 53% in the first two weeks of May, compared to the same weeks in 2019. Although total volume remains low compared to 2019, recent data shows an increase in absolute volume. Using the week of Mar 30 to Apr 5 as a baseline (the period with the lowest observed traffic volume), a 52% average increase in vehicular traffic via MTA bridges and tunnels was observed in May. Average traffic speeds from 8AM to 6PM on north-south avenues between 34th and 57th Streets in Midtown Manhattan remained high (77% higher in May than in Feb 2020), and there was also a 73% increase in school zone speeding tickets from Apr 20 to May 17 as compared to Jan 1 to Mar 12 of 2020. As of May 15, Weigh-in-Motion (WIM) data from C2SMART's testbed on the Brooklyn-Queens Expressway (BQE) shows that the number of very heavy trucks (GVW > 100 kips) remained down 29% for Queensbound (QB) and 44% for Staten Island-bound (SIB) traffic, compared with Feb 2020.

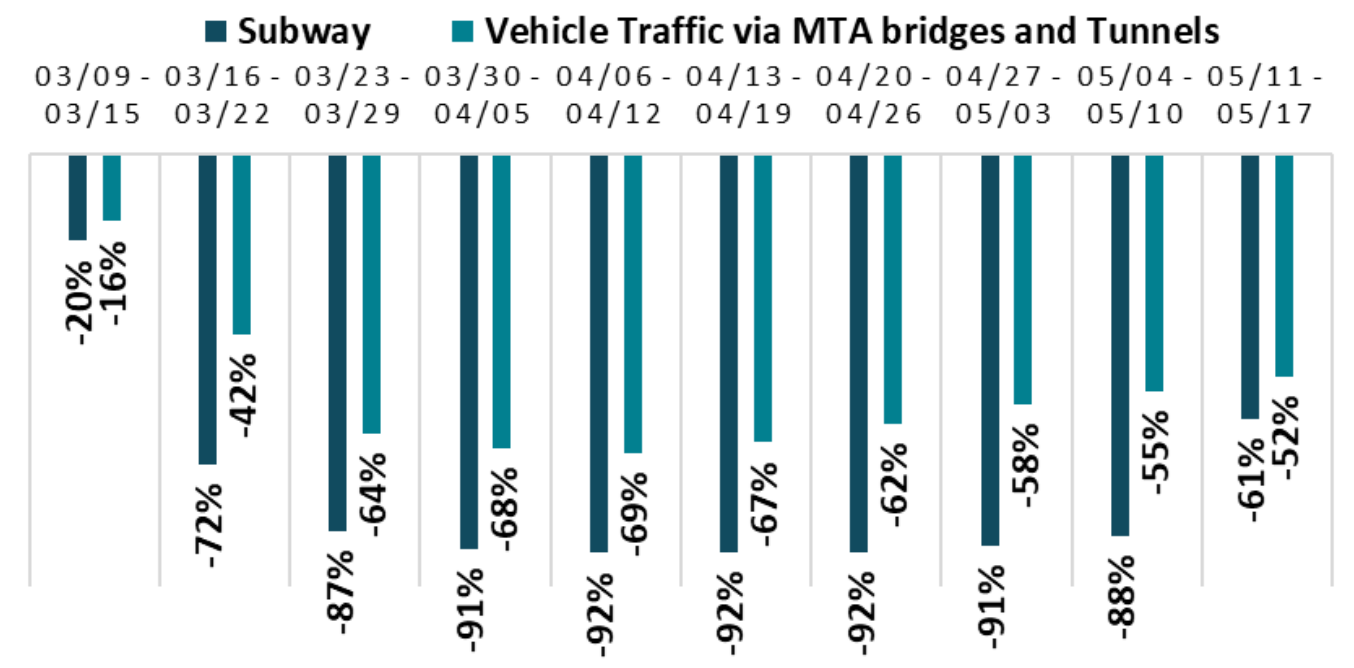

**Figure 1** Mobility Trends in NYC

Seattle, WA saw a 35% increase in I-5 traffic volume in the week of May 25 compared with the week of Mar 30-Apr 5. However, despite a return of vehicular traffic volumes, public transit ridership continues to stay low (-79% in both Apr and May 2020 compared with 2019). Despite the increased vehicular volumes, no significant changes in average travel times were observed on the 495 Corridor in NYC or I-5 in Seattle in May compared with Apr 2020.

## Agent-based Simulation Model to Predict the Impact of COVID-19 Reopening Strategies

Well-calibrated agent-based simulation models built in MATSim (1) were used to study the impact of COVID-19 on the NYC transportation system as well as potential policies for the reopening of NYC. Two baseline models were developed and calibrated: 1) A Pre-COVID model that simulates typical travel behavior, and 2) a COVID model that represents travel behavior during the first month of the COVID-19 pandemic. C2SMART researchers developed the Pre-COVID model called MATSim-NYC (3,5), using a synthetic population of more than 8 million New Yorkers and calibrated transit schedules (4). This open-source, large-scale transportation model covers the entire NYC area and integrates data and modeling capabilities for new technologies and modes. The calibrated models are being used to study:

- Scenarios of multiple reopening phases
- Mode and system usage for various reopening policies and social distancing strategies
- Emissions and air quality impacts

### Pre-COVID Model

*Synthetic Population Data*

The American Community Survey, 2016 Longitudinal Employer-Household Dynamics, and 2040 Socioeconomic and Demographic Forecasts were used to generate personal and household attributes of the synthetic population. The 2010/2011 Regional Household Travel Survey (RHTS) data was employed to prepare travel agendas and model mode choice. To incorporate emerging modes (bike-sharing and ride-hailing), 2016 trip count data of Citi Bike and For-Hire-Vehicles (FHV) were also adopted. The 2017 Citywide Mobility Survey data was used to validate the city-level mode share of the synthetic population.

*Mode Choice Model*

Using travel survey data from the 2010/2011 RHTS from the New York Metropolitan Transportation Council (NYMTC), a mode choice model was proposed to determine mode choice at the tour level in the Pre-COVID model. In addition, the model also incorporated alternatives for emerging mobility services like bike-sharing (e.g., Citi Bike) and ride-hailing (e.g., Uber/Lyft) using observed trip data. Since the default MATSim only supports Multinomial Logit (MNL) model for mode choice, the nested logit model was adjusted to an equivalent trip-level MNL model.

*MATSim Model Network*

The input network was developed with a road network transformed from Open Street Map (OSM) data and a transit network and schedule generated from General Transit Feed Specification (GTFS) data (8,9). For details on model calibration, please see He et al. (4). The network is shown in Figure 2, the layer in green shows the distribution of transit stations.

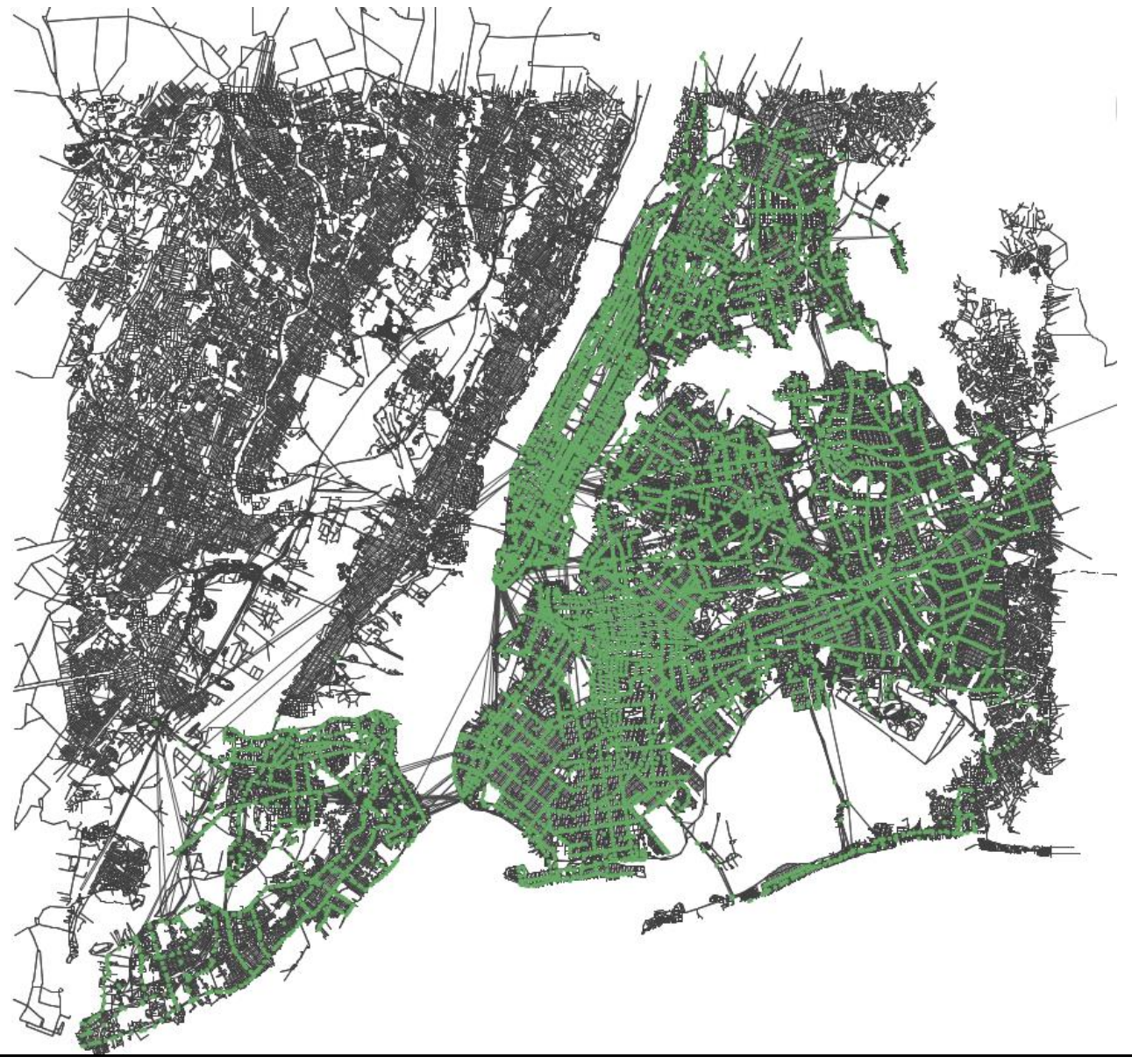

**Figure 2** The road network and transit stations (green)

### COVID Model

The MATSim-NYC model was then adapted and re-fitted to data collected during the pandemic to produce a MATSim-NYC-COVID model.

*MTA Turnstile Data and Apple Mobility Trends Report*

Subway ridership and vehicular traffic data on Metropolitan Transportation Authority (MTA)'s facilities were used for the COVID model. Subway ridership rates of decline reached 89% on average from Mar 23 to Apr 19. This data was supplemented by the Apple Mobility Trends report (7) that reflects requests for directions in Apple Maps. According to Apple data from Mar 23 to Apr 19, trips by transit, driving, and walking decreased by 86%, 58%, and 76%, respectively.

*Work from Home (WFH) Rate*

Determining the WFH rate was crucial in the COVID model development. The classification of working from home for all occupations, based on a study by Dingel and Neiman (2), was merged with this classification with travel demand from the Pre-COVID model based on occupational employment counts for NYC. A 44% WFH rate was calculated in the COVID model, which is very close to the result (42%) found by Dingel and Neiman (2). Figure 3 shows the percentage difference of number of agents generated before/after COVID-19; the result is aggregated by travelers' home Traffic Analysis Zone (TAZ).

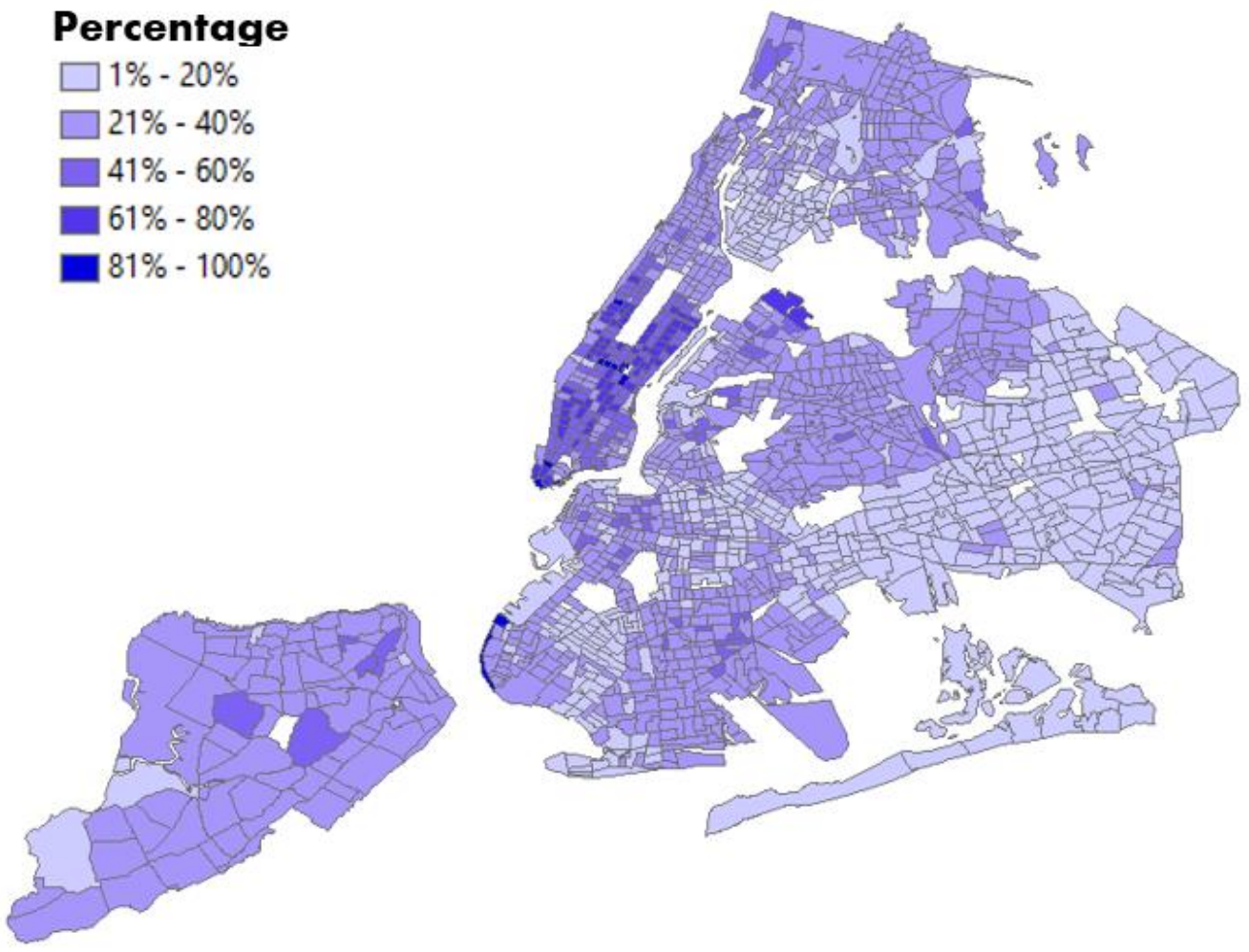

**Figure 3** The difference of number of agents generated before/after COVID-19 based on traveler's home TAZ

The COVID model requires recalibrating the mode choice model to fit the new behavioral setting. The alternative-specific constants for car, transit, walk, and bike were perturbed from the pre-pandemic base model to fit the MTA subway data and Apple Mobility Trends data (6,7). Figure 4 shows the changes of mode share in the Pre-COVID model and in the COVID model. Compared to the Pre-COVID period, the mode share of transit has decreased by 19 percentage points, the mode share of car has increased by 6 percentage points. While these may be considered as extreme-bounded conditions, they reflect general attitudes around contact risk in shared use modes.

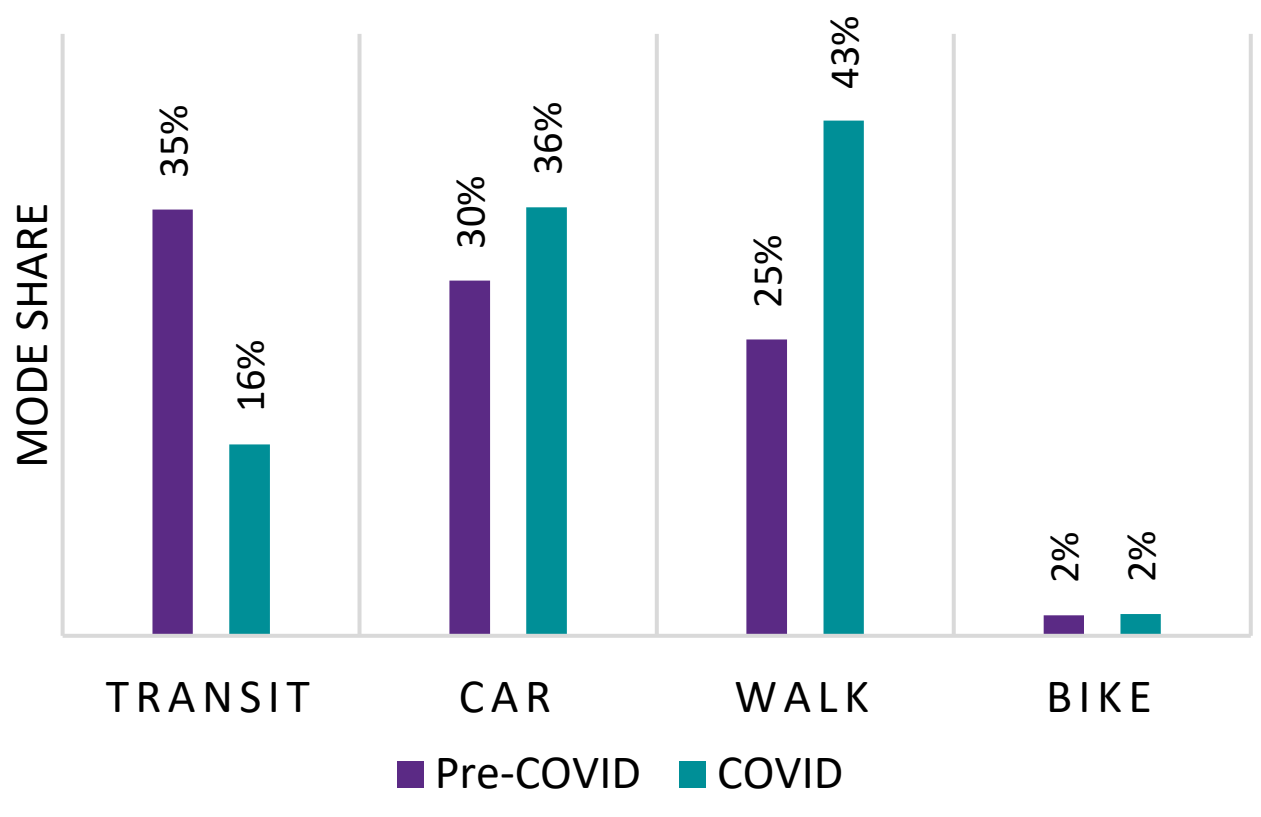

**Figure 4** Changes of mode share in the Pre-COVID model and in the COVID model

The simulation results show trip reductions during the pandemic for subway, car, and walking of 94%, 76%, and 68%, respectively. The average difference between these results and data from MTA turnstile data and Apple mobility trends is about 9%.

## Analyzing Reopening Scenarios

New York State is planning a four-phase reopening based on the regional guidelines (10). Based on these guidelines and the study on occupational work from home rates by study by Dingel and Neiman (2), the corresponding percentage of employees who will return to work was generated. For details on industry related results, please see He et al. (4). A key assumption in these simulation results is that mode preference during the reopening phases will mirror what was observed during the pandemic period due to behavioral inertia. In other words, the following two scenarios can be considered as worst-case for transit based on mode preference not changing from pandemic levels. This assumption might be very conservative and more sensitivity analysis with different assumptions will be tested in the future as more data is revealed.

For the first two reopening phases, the transit schedule was assumed to be the same as it was during the COVID period (using the General Transit Feed Specification (GTFS) on Mar 20, 2020). When simulating the last two reopening phases, the regular transit schedule (GTFS on Jan 20, 2020) was applied. The transit schedule from MTA is continuously changing and the model will be modified as new information is released.

### Scenario 1: Without Any Transit Capacity Restriction

In this scenario, there is no transit capacity restriction, although people are assumed to maintain the same mode preference in the reopening phases as those during the COVID-19 period. Figure 5 shows the changes of mode share in scenario 1 during Pre-COVID period, COVID period and the four reopen phases. Compared with the Pre-COVID model, car mode share increased the most among all the modes tested in the model. The results in Figure 6 show that due to the behavioral inertia, a reopening that does not restrict transit capacity would still potentially only operate at 73% transit ridership from pre-COVID while increasing car traffic to 142% level. In addition, both the number of walking and bike trips are expected to increase in Phase 4. The number of walking trips will be 101% of Pre-COVID levels and bike trips will be 104% of Pre-COVID levels. The mode share in Phase 4 of scenario 1 is shown in Figure 7, showing that the mode share of transit decreased 9 percentage points compared to the pre-COVID period (35%), while the mode share of car increased by 12 percentage points, which was 30% in the Pre-COVID model.

### Scenario 2: 50% Transit Capacity Restriction

Transit agencies around the world are implementing strategies that limit the number of passengers on vehicles to keep riders and workers safe as restrictions on business activity and travel are lifted. For example, the subway system in Beijing, China, which has already been in the reopening stage, limits subway occupancy below 50% of maximum capacity (11). In New Jersey (NJ), NJ Transit trains and buses will operate at 50% capacity as part of orders to maintain social distancing (12). Despite the losses in efficiency, this could be a feasible solution to reduce contact risk and encourage people to use transit.

For Scenario 2, a 50% capacity restriction is applied to all transit vehicles. Again, mode preferences in the COVID-19 period are maintained in the reopening assuming behavioral inertia. The results in Figure 6 show that by Phase 4 only 64% of Pre-COVID transit ridership is restored with the capacity constraint, compared to the 73% transit ridership restoration estimated in Scenario 1. The increase in the number of car trips is around the same in both scenarios (142% of the Pre-COVID model in Scenario 1 and 143% of the Pre-COVID model in Scenario 2). The increase of bike trips is more significant in scenario 2, which is 123% of Pre-COVID model compared to 104% in scenario 1. This shows that with 50% transit capacity restriction, people may prefer to use bikes as an alternative of transit. The mode share in Phase 4 of scenario 2 is shown in Figure 7, the mode share of transit would decrease 12 percentage points compared to the Pre-COVID period (35%), which is 3 percentage points lower than in Scenario 1. The mode share of car would increase 13 percentage points compared to the Pre-COVID period (30%), which is slightly higher than in Scenario 1.

The open-source, modular nature of the MATSim Virtual Testbed has allowed C2SMART researchers to add timely new simulation extensions. Besides the effects of the pandemic and an ensuing recovery on transit use, air quality and emission impact estimation during the COVID-19 pandemic and reopening are being conducted in collaboration with CTECH lab at Cornell University. The current model will be continuously enhanced when more data becomes available.

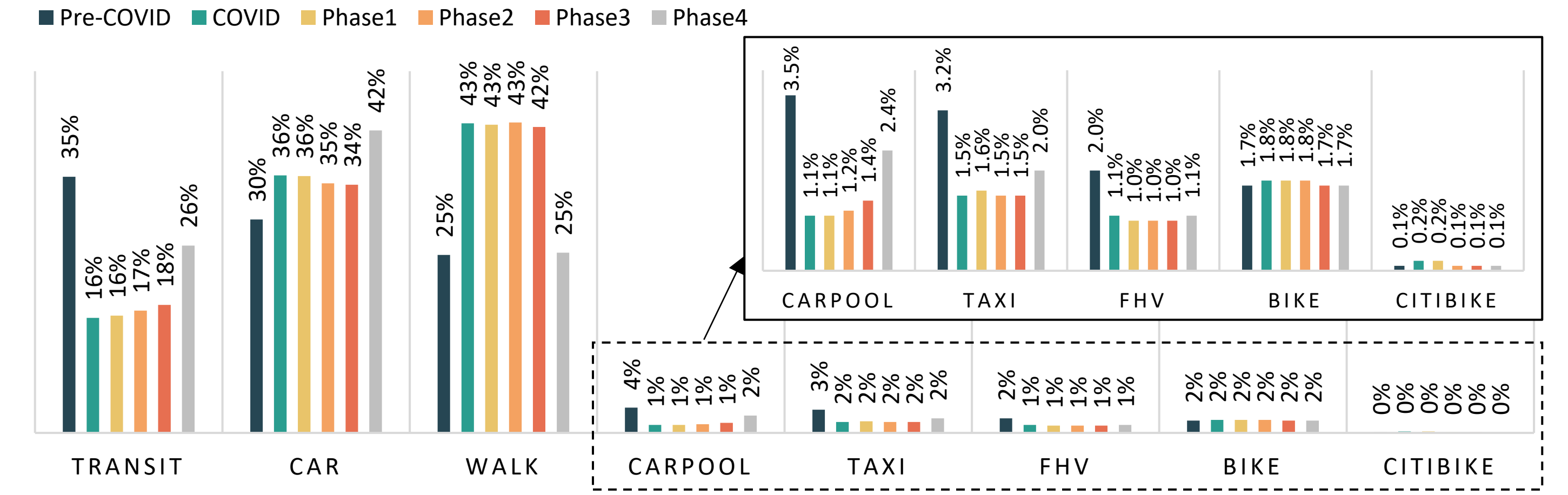

**Figure 5** Mode share prediction in the Pre-COVID 19 model, COVID model and Reopening phases (scenario 1)

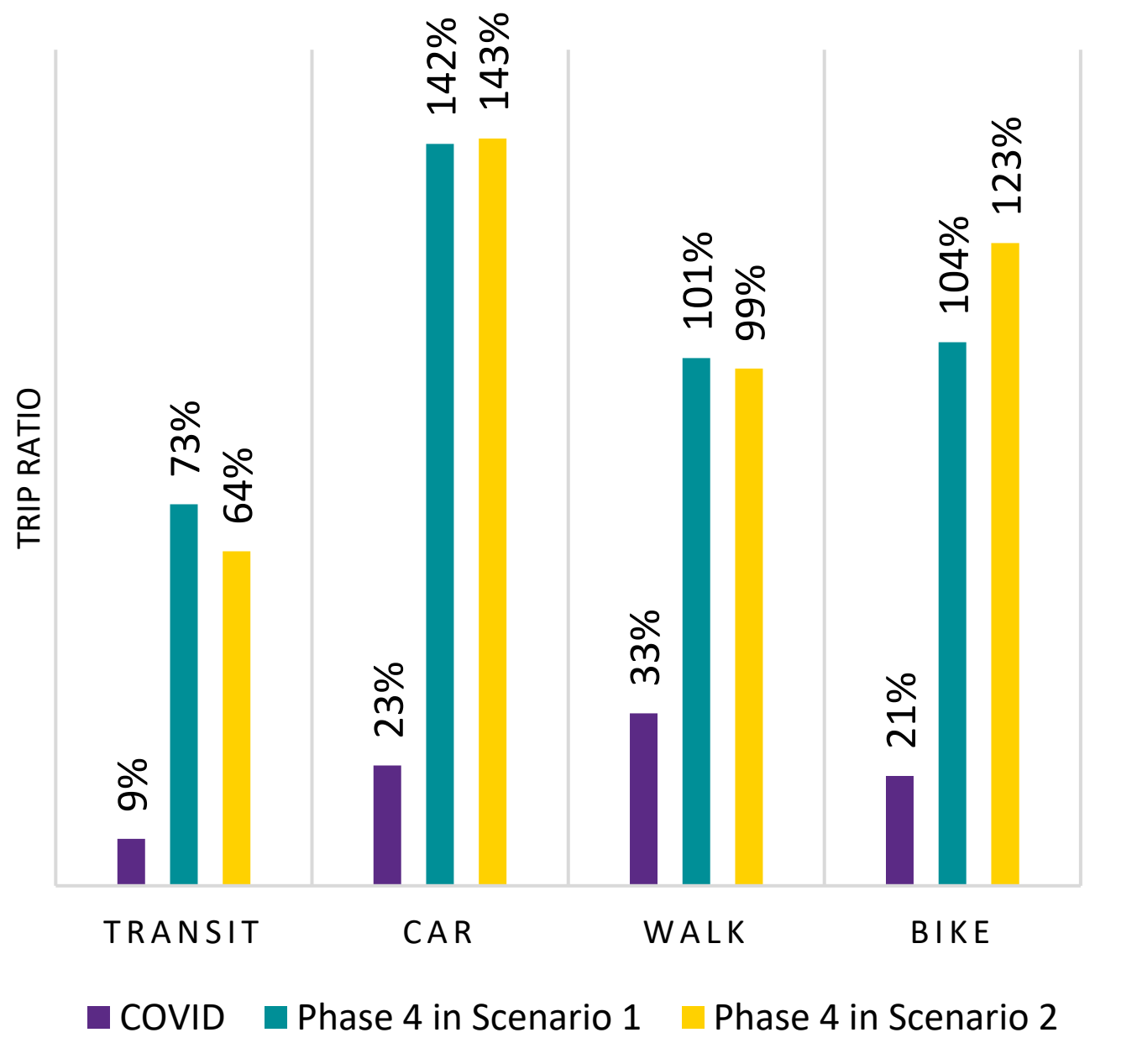

**Figure 6** The trip ratio in the COVID model and Phase 4 in two simulation scenarios (with and without transit capacity restriction) compared to the Pre-COVID model

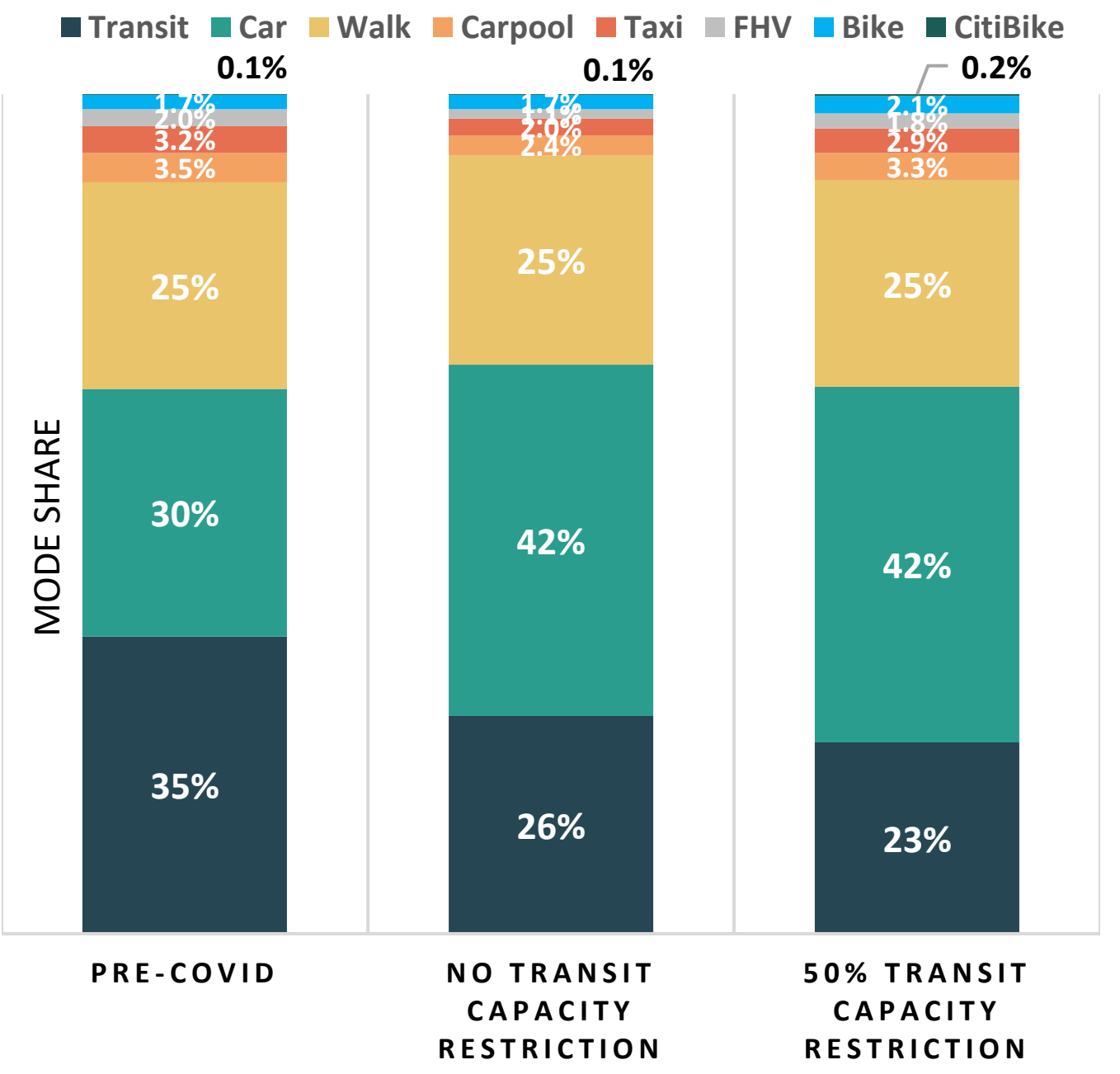

**Figure 7** The mode share comparison in the Pre-COVID model and Phase 4 in two simulation scenarios (with and without transit capacity restriction)

## Sociability Indicators from Real-time Street Cameras

According to the World Health Organization and the CDC, social distancing is currently the most effective way to slow the spread of COVID-19. Although social distancing orders are mandated, data on how people are responding to these policies is not widely available. Understanding the actual reduction in social contact is important to measuring the effectiveness of the policy, especially as the crisis begins to ease and volumes of people traveling within the city start to increase. The C2SMART research team developed a novel social-distancing dataset along from a video processing method based on deep learning for analyzing time-dependent, social-distancing, patterns at the local level in the COVID-19 pandemic and the subsequent recovery process. Object detection was applied to real-time traffic camera videos (13) at multiple key locations within NYC and Seattle to provide information about crowd density. One of the state-of-the-art algorithms, RetinaNet (14), and ResNet-50 (15) are used as the backbone network architecture. The model was pre-trained using the COCO dataset (16). The open-source platforms TensorFlow (17) are utilized as the machine learning library's support, along with OpenCV (18) for video processing. Figure 8 presents the video processing and data analysis process.

To calculate the social distancing safety rate (number of pedestrian pairs > 6 feet proximity/total number of pedestrian pairs), the centroids of detected pedestrians are identified and the distance between the centroids are calculated. Next, the ratio of real height and pixel height (R-P ratio) are computed to project the real distance between people by assuming every person has the same height (1.70 meters/5.58 feet is used).

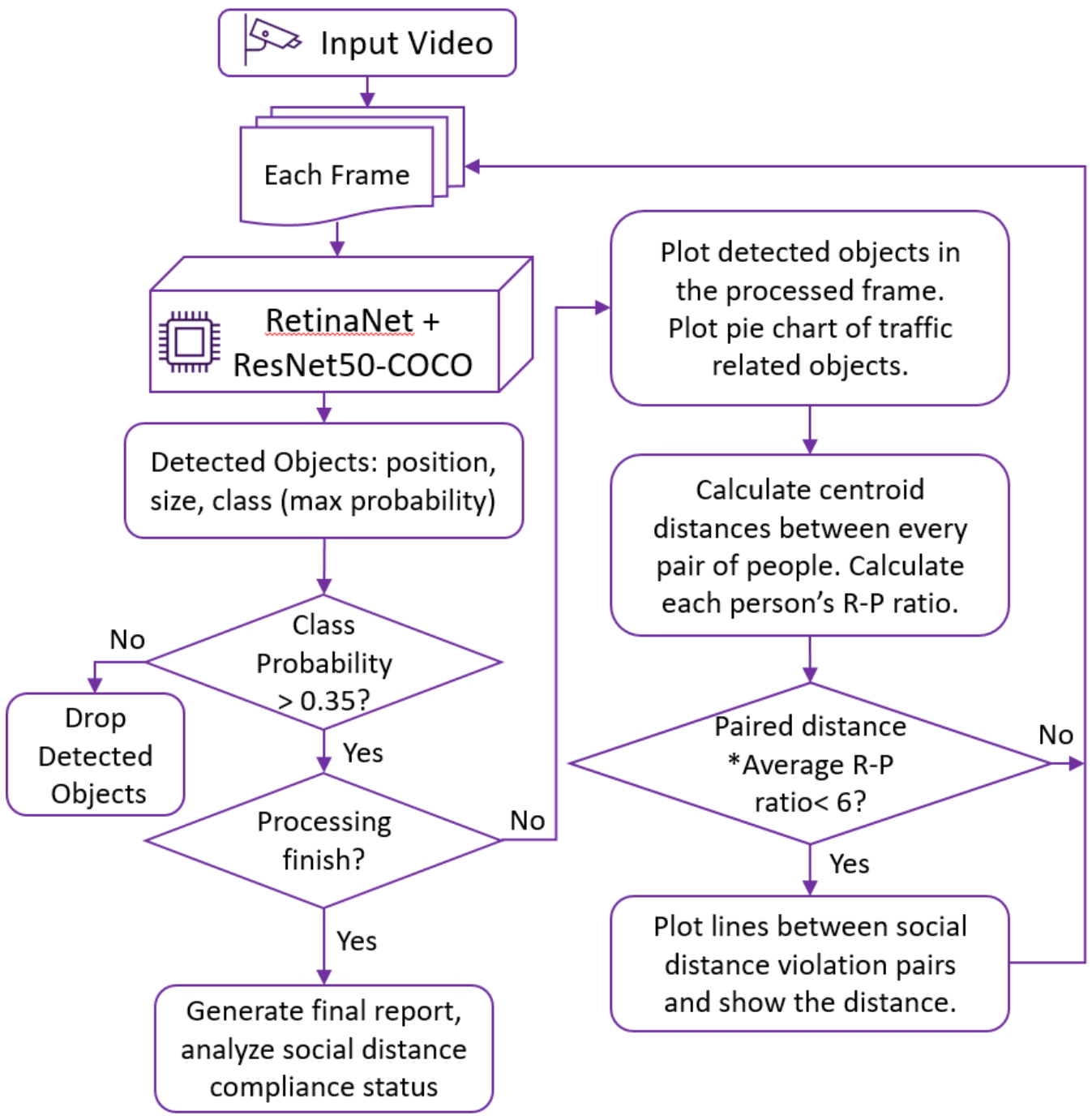

**Figure 8** Video processing and data analysis process

Traffic-related objects (person, car, truck, bicycle, bus) and the total number of social distancing pairs less than 6 feet proximity are reported for each frame. Figure 9 presents an example of the video processing output around the Main Street and Roosevelt Avenue intersection in Flushing, Queens and the preliminary sociability metrics based on 7 key locations in NYC are summarized in Table 1. A gradual increase in pedestrian volumes is observed, especially in the last two weeks of May. Accordingly, the social distancing safety rate dropped slightly. The relationship between pedestrian density and social distancing safety rate is also examined and shown in Figure 10.

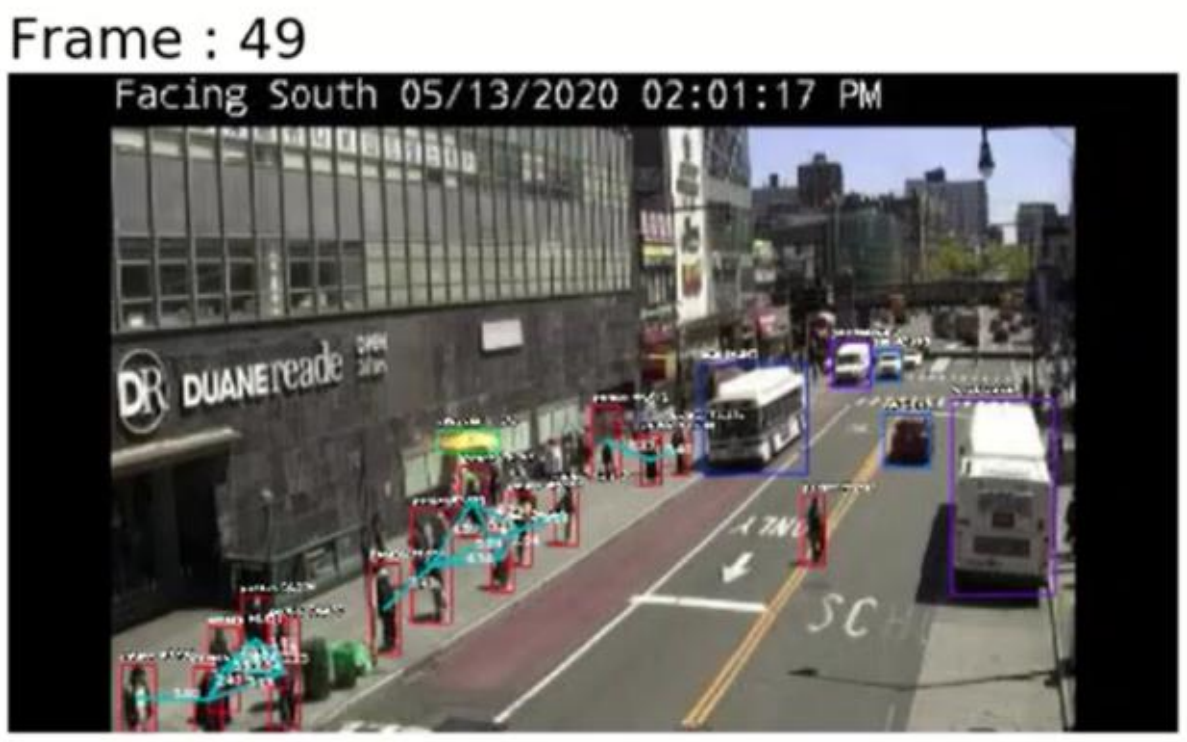

(a) Object detection (blue lines highlight the pedestrian pairs whose distance is less than 6 feet)

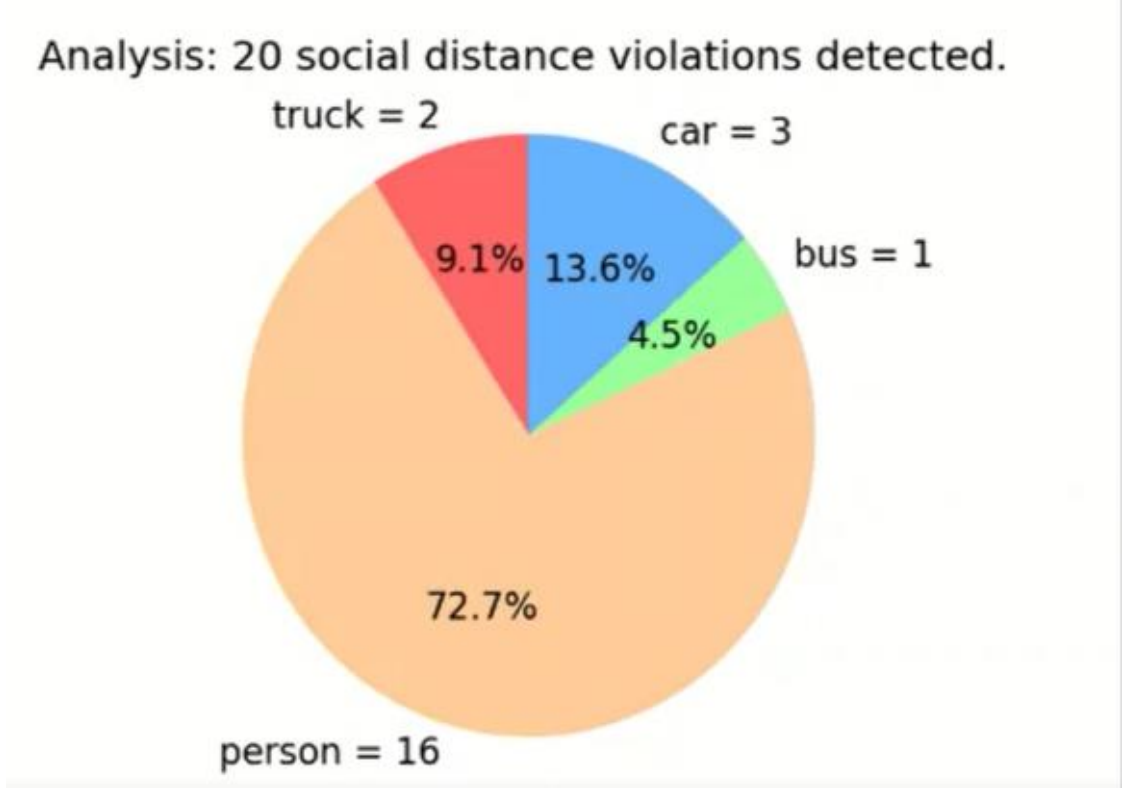

(b) Crowd density pie chart

**Figure 9** Example of video processing output (Main St and Roosevelt Avenue, Queens, NYC)

The social distancing detection algorithm used in NYC was also implemented for a local street intersection (Broadway & E Pike St EW) in Seattle with a frame rate of 30 seconds. Results from May 18 and June 1 show a similar average and maximum pedestrian density. However, the pedestrian social distancing safety rate is reduced from 88% to 85%. More camera data will be collected and enlarge the coverage of social distancing safety detection. Additional locations and examples are available at c2smart.engineering.nyu.edu/covid-19-dashboard.

**Table 1 COVID-19 Sociability Metrics (NYC)**

| **NYC, NY** | Apr 2 | May 13 | May 27 |
|---|---|---|---|
| Average Peds Density (#/frame) | 2.6 | 2.9 | 3.5 |
| Maximum Peds Density (#/frame) | 20 | 24 | 19 |
| Social Distancing Safety Rate | 91% | 87% | 86% |
| **Seattle, WA** | | May 18 | Jun 1 |
| Average Peds Density (#/frame) | | 3.2 | 3.0 |
| Maximum Peds Density (#/frame) | | 12 | 11 |
| Social Distancing Safety Rate | | 88% | 85% |

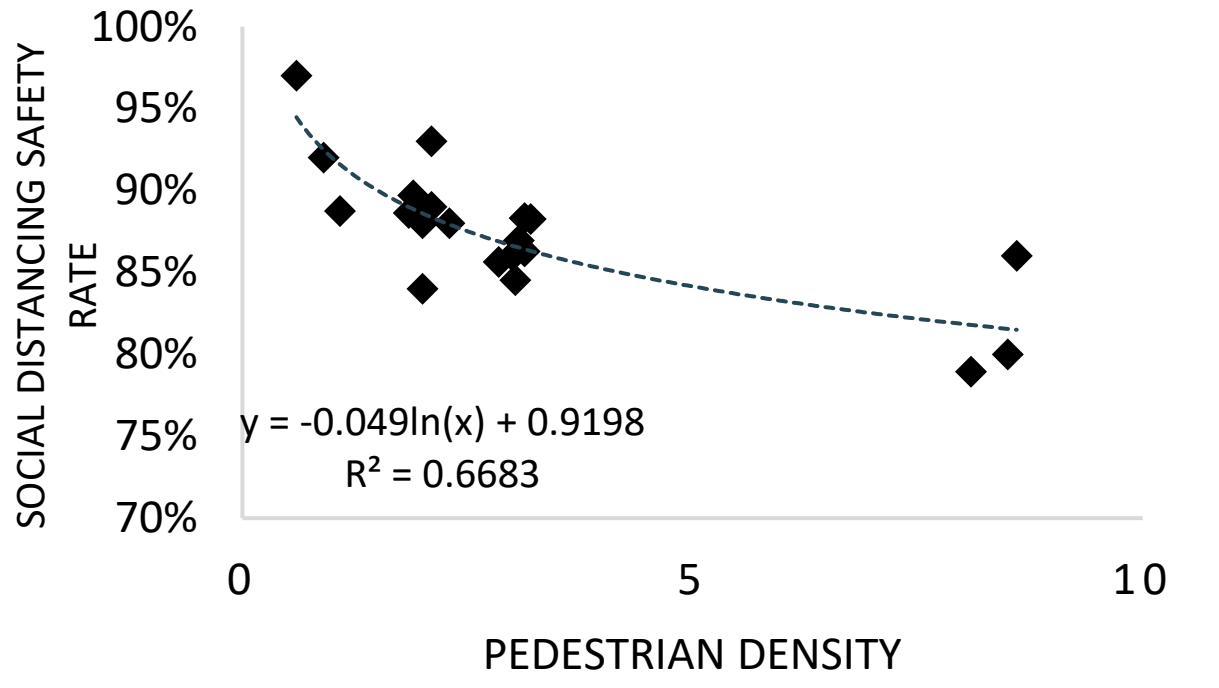

**Figure 10** Pedestrian density and social distancing safety rate

## Temporal Density Distribution

Based on the objective detection outputs from the cameras, temporal density distribution profiles are constructed for pre-pandemic conditions to investigate potential temporal pattern changes. Figure 11 shows the 24-hour distributions for pedestrian, car, and cyclist density at 5th Avenue and 42nd Street, Manhattan, NYC. This location is typically a high-density area and has an overall high and consistent pedestrian density through the day (Figure 11 (a) navy line) before the crisis. While car and pedestrian density remain low in Apr and May 2020 compared with pre-pandemic levels, the emergence of commuter spikes (morning and afternoon peak for pedestrians and morning peak for cars) was observed. One possible reason could be that most of the observed pedestrians are essential workers. Unlike pedestrian and car density, cyclist density has almost approached pre-pandemic level at this location by the end of May.

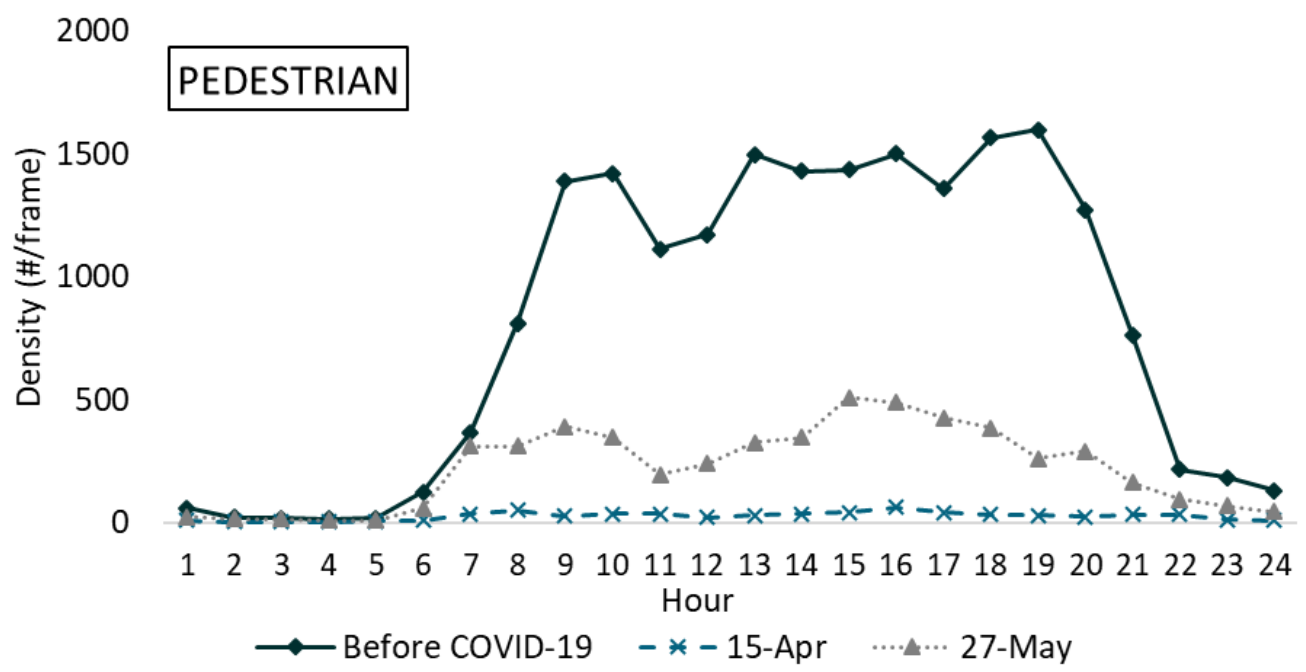

(a) Pedestrian density distribution

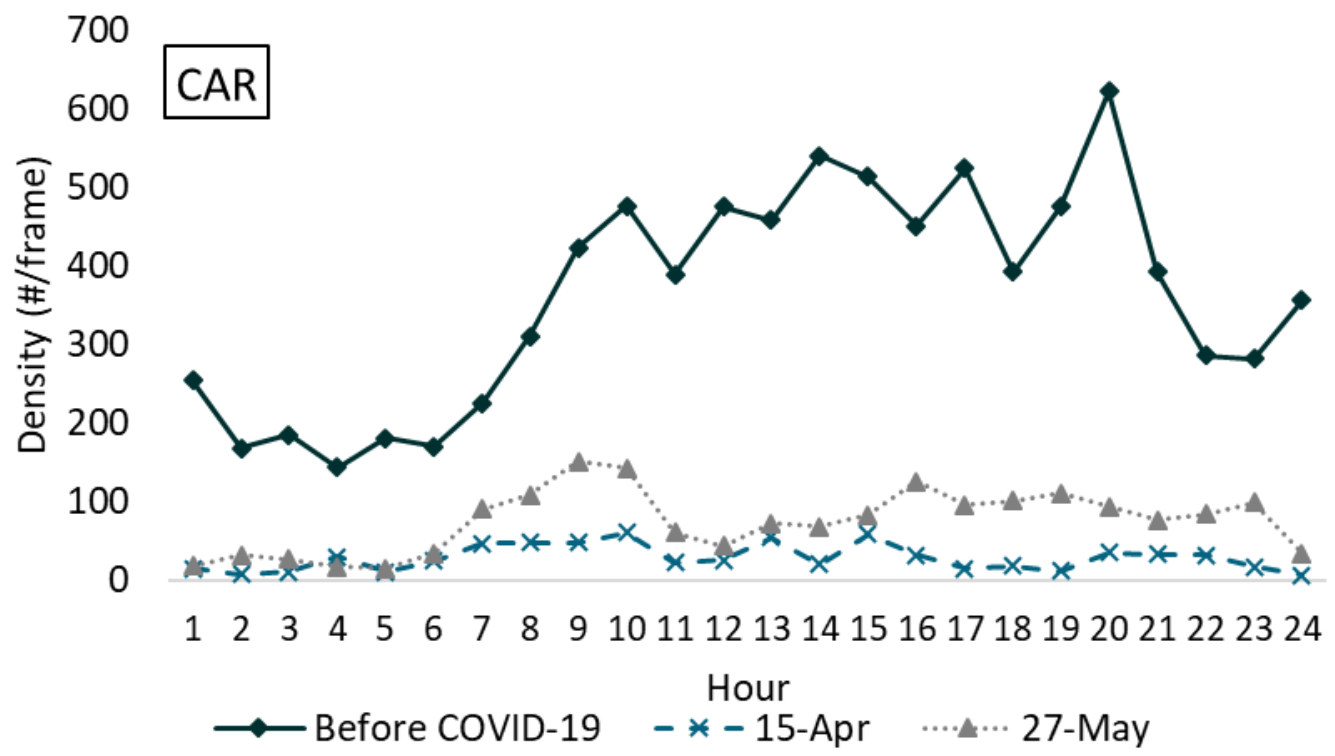

(b) Car density distribution

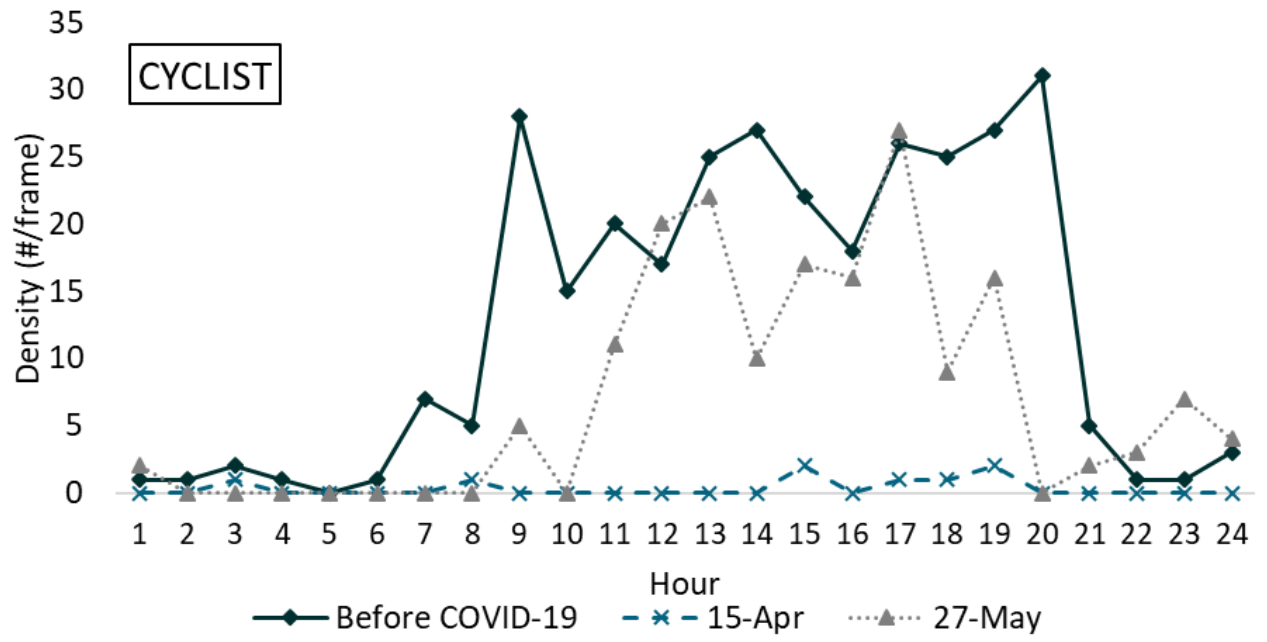

(c) Cyclist density distribution

**Figure 11** Temporal Distributions of Pedestrian, Car and Cyclist Density (5 Ave/42 St, Manhattan)

As an on-going effort, this approach will be extended to cover 100 camera locations to continuously evaluate the changes in crowd density and social distancing practice between pedestrians. With cloud computing power, it is also possible to translate the findings to actionable operational items in near real-time to track the density trends during the reopening phases and further utilized for predictive analysis (e.g. future cycling rates), to assist developing effective strategies or to plan for potential future scenarios.

## Summary of Findings

As NYC begins the first phase of reopening, many questions and challenges for transportation systems remain. C2SMART researchers recalibrated their simulation testbed to evaluate the impact of COVID-19 on travel behavior and estimate possible mode share changes for different reopening phases with the consideration of social distancing.

In addition, crowd density and social distancing for pedestrians based on data obtained from publicly available camera feeds and using state-of-the-art video processing techniques can provide useful insights into pedestrian and cyclist volumes and their behavior in terms of maintaining social distance.

This paper reflects the Center's perspective as of June 11, 2020 based on data collected in May 2020. C2SMART researchers are continuing to collect data and monitor both the mobility and sociability trends and regularly update findings after the reopening of the city.

*Note: all data is preliminary and subject to change.*

*For more information please contact* c2smart@nyu.edu

*For more data and visualizations, please visit our COVID-19 Dashboard:* c2smart.engineering.nyu.edu/covid-19-dashboard/